\documentclass[english,twocolumn,amssymb,amsmath,superscriptaddress]{revtex4}
\usepackage[latin9]{inputenc}
\usepackage{graphicx,color}
\usepackage{hyperref}
\usepackage{babel}
\usepackage{booktabs}
\usepackage[normalem]{ulem}

\newcommand{\be}{\begin{equation}}
\newcommand{\ee}{\end{equation}}

\definecolor{mygreen}{rgb}{0,0.5,0}
\definecolor{myblue}{rgb}{0,0,0.75}
\definecolor{mymagenta}{cmyk}{0,1,0,0.12}

\def\ket#1{\left| #1\right\rangle}  

\begin{document}

\title{Chiral Topological Phases from Artificial Neural Networks}

\author{Raphael Kaubruegger}
\affiliation{Department of Physics, University of Gothenburg, SE 412 96 Gothenburg, Sweden}
\affiliation{Institute for Theoretical Physics, University of Innsbruck, A-6020 Innsbruck, Austria}
\author{Lorenzo Pastori}
\affiliation{Department of Physics, University of Gothenburg, SE 412 96 Gothenburg, Sweden}
\affiliation{Institute of Theoretical Physics, Technische Universit\"at Dresden, 01062 Dresden, Germany}
\author{Jan Carl Budich}
\affiliation{Department of Physics, University of Gothenburg, SE 412 96 Gothenburg, Sweden}
\affiliation{Institute of Theoretical Physics, Technische Universit\"at Dresden, 01062 Dresden, Germany}	
\date{\today}

\begin{abstract}
Motivated by recent progress in applying techniques from the field of artificial neural networks (ANNs) to quantum many-body physics, we investigate as to what extent the flexibility of ANNs can be used to efficiently study systems that host chiral topological phases such as fractional quantum Hall (FQH) phases. With benchmark examples, we demonstrate that training ANNs of restricted Boltzmann machine type in the framework of variational Monte Carlo can numerically solve FQH problems to good approximation. Furthermore, we show by explicit construction how $n$-body correlations can be kept at an exact level with ANN wavefunctions exhibiting polynomial scaling with power $n$ in system size. Using this construction, we analytically represent the paradigmatic Laughlin wavefunction as an ANN state.   

\end{abstract}

\date{\today}

\maketitle

\section{Introduction }
The quest for methods to solve, at least approximately, the quantum many-body problem has been a major focus of research in physics for many years. The paramount issue in this context is the exponential complexity of the wavefunction, which severely limits the system sizes tractable with exact diagonalization. An important challenge for the study of larger quantum many-body systems is to efficiently parameterize the physically relevant states. Along these lines, Carleo and Troyer \cite{CarleoTroyer} recently demonstrated the potential of artificial neural networks (ANNs) as an ansatz for variational wavefunctions. There, the synaptic coupling strengths between the physical (visible) and auxiliary (hidden) spin variables (neurons) of the ANN play the role of the variational parameters, and the quantum state is obtained by tracing out the auxiliary variables.   

The purpose of this work is to harness the flexibility of ANNs to study chiral topological phases (CTPs) in two spatial dimensions (2D), such as fractional quantum Hall states \cite{StormerFQH,LaughlinWave,PrangeGirvinBook} and chiral spin liquids \cite{AndersonQSL,KalmeyerLaughlin,WenBook} in the framework of variational Monte Carlo (VMC). Furthermore, we analytically demonstrate how CTP model wavefunctions can be exactly represented with ANNs {\emph{at polynomial cost}}. This is of particular relevance as these exotic phases so far have quite obstinately eluded efficient numerical methods: Quantum Monte Carlo approaches to finding CTP ground states are generically stymied by the negative sign problem, and fundamental limitations regarding the exact representability of such complex many-body states with tensor networks have been proven \cite{DubailRead2015,Read2017}. However, despite these difficulties, it is fair to say that impressive progress has been made in the computational treatment of CTPs, e.g. using matrix product states at the expense of exponential scaling of resources in only one of the spatial directions \cite{WhiteDMRG,McCullochiDMRG,ZaletelExactMPS}, and tree-tensor network methods \cite{VidalTTN, NoackTTN, MarcelloSimoneTTN}. Another promising direction is to resort to tensor network states of mixed state density matrices the effective temperature of which decreases with increasing resources \cite{BeriCooperMixed}. Furthermore, Monte Carlo techniques have been successfully applied using e.g. the fixed phase method \cite{Ortiz93,Ortiz97}, and to sample various sample various observables from CTP model wavefunctions (see e.g. \cite{Haldane2017_LatticeMC} and references therein).

\begin{figure}[h!]
\centering
\includegraphics[width=0.95\columnwidth]{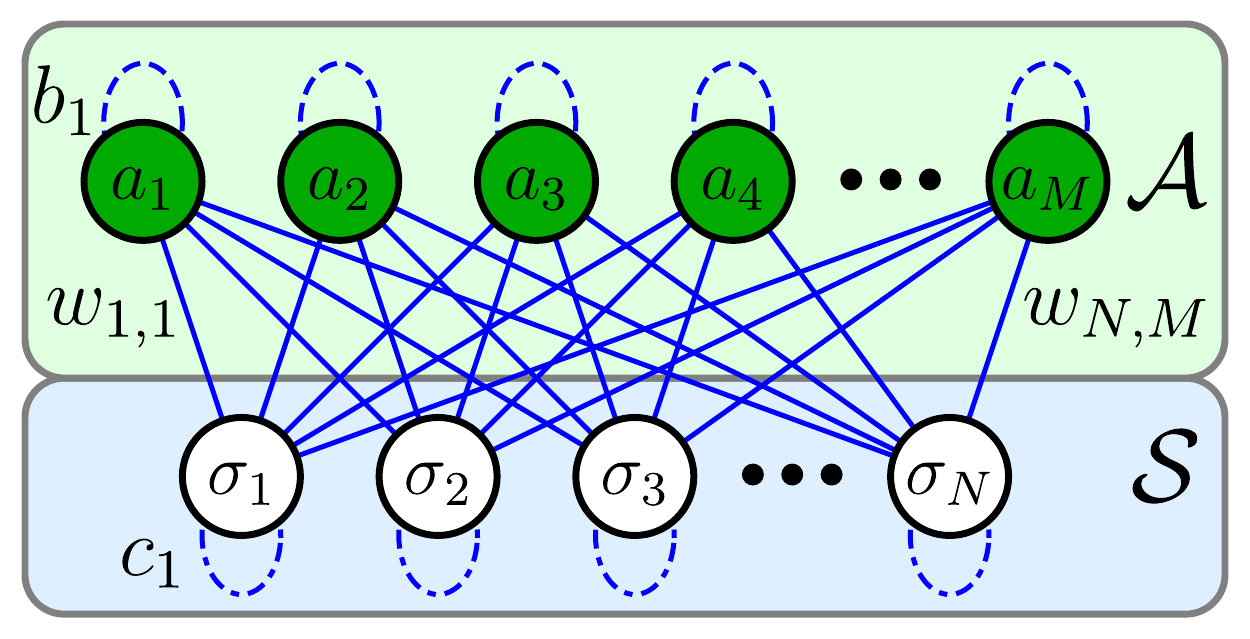}
\caption{(color online) Graphical representation of the restricted Boltzmann machine (RBM) network for the variational wavefunction [see Eq.~(\ref{eqn:generalpsi})]. The physical spins (in the blue shaded area $\mathcal S$) are denoted by $\sigma_j,~ j=1,\ldots, N$, and the auxiliary variables (in the green shaded area $\mathcal A$) are denoted by green dots $a_{j},~j=1,\ldots, M$ with $M=\alpha N$. The coupling strengths on the links between $i$ and $j$ (solid lines) are labeled $w_{ij}$, while the local fields are denoted by $b_i$ (dashed lines) for the auxiliary variables and by $c_j$ (dashed-dotted lines) for the physical spins, respectively.}
\label{fig:one}
\end{figure}

\begin{figure}[htp]
\centering
\includegraphics[width=0.95\columnwidth]{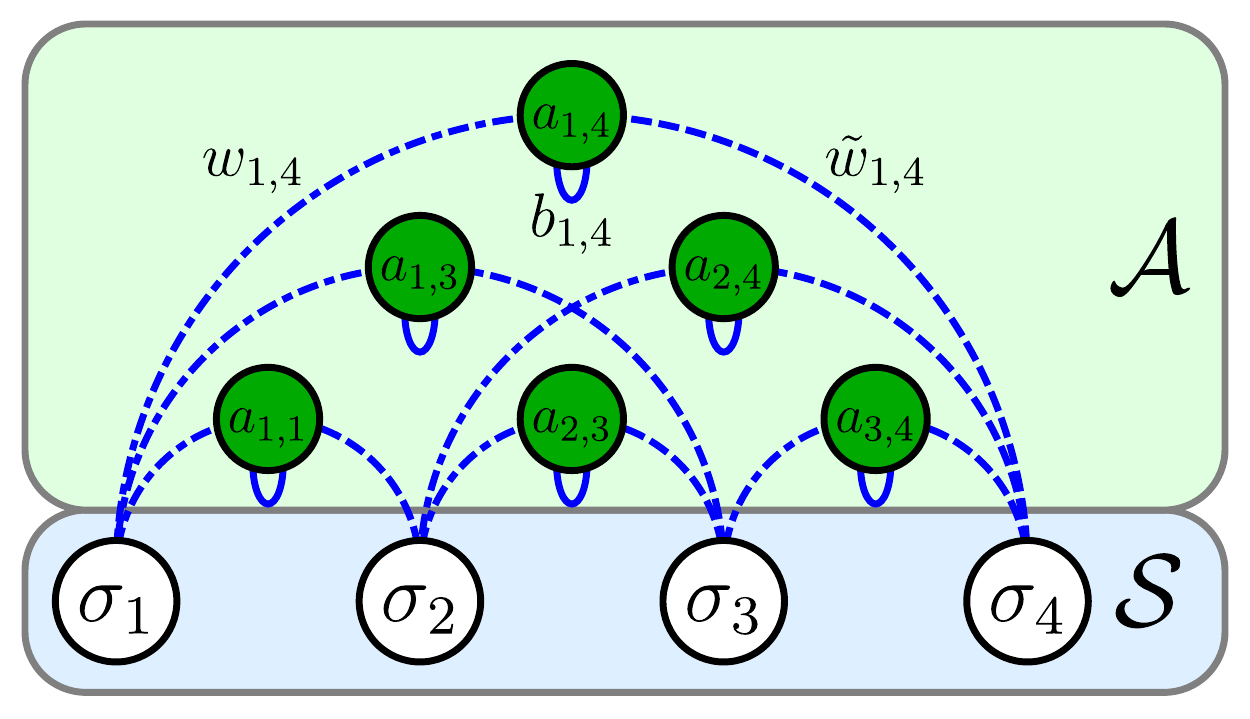}
\caption{(color online) Sketch of the cluster neural network quantum state (CNQS) architecture with cluster size $n=2$ and $m=1$ [see Eq.~(\ref{eqn:generalpsi2})]. The physical spins (in the blue shaded area $\mathcal S$) are denoted by $\sigma_i$. The auxiliary variables (in the green shaded area $\mathcal A$) are denoted by $a_{ij}$. The coupling strengths on the links between $i\ne j$ (dashed-dotted lines) are labeled $w_{ij}$ and $\tilde w_{ij}$, respectively, while the local fields (solid lines) at $a_{ij}$ are labeled $b_{ij}$. For $n>2$ (not shown), $m$ auxiliary variables $a_{i_1,\ldots i_n}^\nu,~\nu=1,\ldots m$ are associated with every cluster of $n$ distinct sites labeled $i_1,\ldots i_n$.}
\label{fig:two}
\end{figure}

Below, we study 2D lattice systems hosting CTPs within the ANN architecture of restricted Boltzmann machines (RBM) \cite{CarleoTroyer} [see Fig.~\ref{fig:one} for an illustration]. Using VMC techniques to train the network, we investigate the efficiency of this method in finding the ground state of chiral spin liquid and lattice fractional quantum Hall models such as the Kapit Mueller model \cite{GreiterParent,KapitMueller}. As a benchmark for small systems, we compare our VMC results to exact diagonalization. Remarkably, we find that systems the size of which exceeds the scope of exact diagonalization can be solved with the ANN approach, by increasing the number of variational parameters {\emph{polynomially}} with system size \cite{footPoly}. Besides this numerical study, we construct a modified RBM architecture, coined cluster neural network quantum states (CNQS) [see Fig.~\ref{fig:two} for an illustration], to capture CTP model states. While many tensor network methods rely on the truncation of entanglement in real space, the CNQS ansatz is based on limiting the number of particles that are directly correlated in the wavefunction as a means to contain its complexity. For example, the Laughlin state as a paradigmatic representative of CTPs is characterized by the constraint of simultaneously maximizing the relative angular momentum between any pair of particles. Such two-body constraints of Jastrow form are exactly captured by a CNQS with quadratic scaling [see Fig.~\ref{fig:two}] as we show analytically. Three body-constraints which appear in non-Abelian phases such as the Moore-Read state \cite{MooreRead} require a CNQS ansatz with cubic effort in system size.

This article is structured as follows. In Section \ref{sec:rbm}, we discuss how variational wavefunctions are obtained from the RBM architecture. Thereafter, in Section \ref{sec:main} we apply this RBM variational ansatz to numerically study chiral topological phases, and introduce the CNQS architecture in Section \ref{sec:cnqs} to obtain analytical insights as to how CTP model states can be exactly described with ANNs. Finally, a concluding discussion is presented in Section \ref{sec:discussion}. Technical details about the numerical methods we use in this work are provided in the appendix. 

\section{Restricted Boltzmann machine states } 
\label{sec:rbm}
The general ANN framework considered here is that of an RBM consisting of a set of $N$ physical spins $\left\{\sigma_1,\ldots,\sigma_N\right\}=\mathcal S$ coupled to a set $\mathcal A$ of $M$ classical Ising spins called the auxiliary (hidden) variables, via a set $\mathcal W$ of complex parameters \cite{CarleoTroyer}. The network energy of the RBM is then defined as $\mathcal{E}_{\text{nw}}(\mathcal S, \mathcal W, \mathcal A)=\sum_{j}\sigma_j c_j +\sum_i(\sum_j w_{ji}\sigma_j + b_i)a_i$, where $w_{ij} \in \mathcal W$ are the couplings between the auxiliary and the physical spins, while the $b_{i}, c_j\in \mathcal W$ play the role of a complex local field for the auxiliary variables $a_i=\pm 1$ and the physical spins $\sigma_j=0,1$, respectively. The network energy  $\mathcal{E}_{\text{nw}}$ does not have the meaning of a physical energy, but  specifies the connectivity of the RBM via the functional form of a Boltzmann weight. The defining constraint of an RBM is that there are no direct couplings within $\mathcal A$ which allows to analytically trace out the auxiliary variables, yielding the explicit form of the variational wavefunction at fixed couplings $\mathcal W$:
\begin{align}
&\psi_{\mathcal W}(\mathcal S)=\sum_{\left\{a_{i}\right\}}\text{e}^{-\mathcal{E}_{\text{nw}}(\mathcal S,\mathcal W,\mathcal A)}=\nonumber\\
&\text{e}^{-\sum_jc_j \sigma_j}\prod_{i}2\cosh(\sum_j w_{ij}\sigma_j + b_{i}).
\label{eqn:generalpsi}
\end{align}
Choosing a constant density $\alpha$ of auxiliary variables per physical spin, i.e. $M=\alpha N$, the number of variational parameters scales as $\alpha N^2$.

\section{Chiral topological phases from RBM states. }
\label{sec:main}
We now demonstrate how the RBM variational wave function [see Eq.~(\ref{eqn:generalpsi})] approach can be used to solve systems hosting CTPs. Concretely, we study the lattice model introduced by Kapit and Mueller \cite{KapitMueller} on a 2D square lattice. Considering the limit of hardcore-bosons, the model Hamiltonian can be readily cast into the spin-1/2 form
\begin{align}
H=\sum_{jk}J_{j,k}S^+_jS^-_k,
\label{eqn:KapitMuellerHam}
\end{align}  
where the spin operators $S^\pm_j=S^x_j\pm iS^y_j$ at site $j=(x_j,y_j)$ replace the bosonic creation and annihilation operators $\hat a_j^\dag$ and $\hat a_j$, respectively. Introducing the complex notation $z_j=x_j+iy_j$ for the 2D lattice indices $j$, the complex coupling matrix elements $J_{j,k}$ take the form \cite{KapitMueller}
$J_{j,k}\,=\,W(z)\,e^{i\pi\phi\,(y_k-y_j)(x_k+x_j)}$ 
with $z=z_k-z_j=x+iy$ and the exponentially decaying prefactor $W(z)$ reads as
$W(z)\,=\,(-1)^{x+y+xy}\exp\left\{-\frac{\pi\,(1-\phi)}{2}\,|z|^2\right\}$, where $\phi$ is the magnetic flux per plaquette. The single particle states of the Kapit-Mueller Hamiltonian constitute a lattice version of the lowest Landau level in the continuum, and the appearance of fractional quantum Hall states as its many-body ground states has been proven in several studies \cite{KapitMueller,LiuBergholtzKapit,EmilReview}. In our present numerical study, we consider a quarter filling of the lattice with hardcore bosons (i.e. $N/4$ spin up sites in the spin language) at flux $\phi=1/2$. At these parameters, a bosonic $\nu=\frac{1}{2}$ Laughlin phase and the corresponding chiral spin liquid phase in the spin language, respectively, are the ground states of this model. Specifically, we consider the Hamiltonian of Eq.~(\ref{eqn:KapitMuellerHam}) in a cylinder geometry, with periodic boundary conditions in $y$ direction. As chiral edge states appear in this geometry, reaching variational energies close to the actual ground state energy implies that also these edge states are well captured by the RBM wavefunction (\ref{eqn:generalpsi}).

We initialize the RBM with a set of random parameters $\mathcal W$, generally using $\alpha=4$, and search for the ground state of the Hamiltonian (\ref{eqn:KapitMuellerHam}) by minimizing the energy expectation value of the RBM state (\ref{eqn:generalpsi}) using the stochastic reconfiguration (SR) method to update the RBM wavefunction \cite{CarleoTroyer, Sorella2001, SorellaSR, MinresQLP,sup}.  In Table \ref{table:one}, we compare the results we obtain from exact diagonalization (ED) to those from the RBM ansatz for various system sizes. For system size $L_x\times L_y=8\times 8$, the Hilbert space dimension after taking into account particle number conservation and translation symmetry is $6.1\times10^{13}$ and thus beyond the scope of direct study with ED. However, for such larger systems we interpolate the expected ground state energy by noticing that the deviation of the ground state energy from $-N/4$ up to small fluctuations only depends on the circumference of the cylinder [see values marked with a $*$ in Table \ref{table:one}]. With our VMC calculations, we reach down to the ground state energy up to a relative deviation $\Delta E_{\text{rel}}$ on the order of $10^{-4}$ to $10^{-3}$, where the difference to the exact energy is found to be least at the smallest circumference $L_y=4$, owing to the smaller influence of the metallic edge effects at longer aspect ratios. 

\begin{table}
	\begin{tabular}{l || c | c | c}
		\toprule[0.1pt]
		Size  &  ED  &  VMC & $\Delta E_{\text{rel}}$ \\
		\midrule[0.5pt]
		$4\times4$  &  $-3.8776$  &  $-3.8769(3)$ & $1.7\times10^{-4}$ \\
		$6\times4$  &  $-5.8773$  &  $-5.8767(3)$ & $1.0\times10^{-4}$ \\
		$8\times4$  &  $-7.8773$  &  $-7.8764(3)$ & $1.1\times10^{-4}$ \\
		$4\times6$  &  $-5.7125$  &  $-5.7019(8)$ & $1.9\times10^{-3}$ \\
		$6\times6$  &  $-8.712^*$ &  $-8.7010(8)$ & $1.3\times10^{-3}$ \\
		$4\times8$  &  $-7.6632$  &  $-7.658(1)$ & $6.7\times10^{-4}$ \\
		$8\times8$  &  $-15.663^*$  & $-15.652(2)$ & $6.9\times10^{-4}$ \\
		\bottomrule[0.1pt]
	\end{tabular} 
\caption{Comparison between the ground state energy of the Kapit-Mueller Hamiltonian [see Eq.~(\ref{eqn:KapitMuellerHam})] in cylinder geometry obtained by exact diagonalization (ED) and variational Monte Carlo (VMC), for different system sizes $L_x\times L_y$. In the ED coulmn, the values marked with $*$ are interpolated from shorter cylinders with the same circumference. The fourth column shows the relative deviation of the VMC result from the ED, defined as $\Delta E_{\text{rel}}=(E_{\text{ED}}-E_{\text{VMC}})/E_{\text{ED}}$.}
\label{table:one}
\end{table}

\begin{figure}[htp]
	\centering
	\includegraphics[width=0.95\columnwidth]{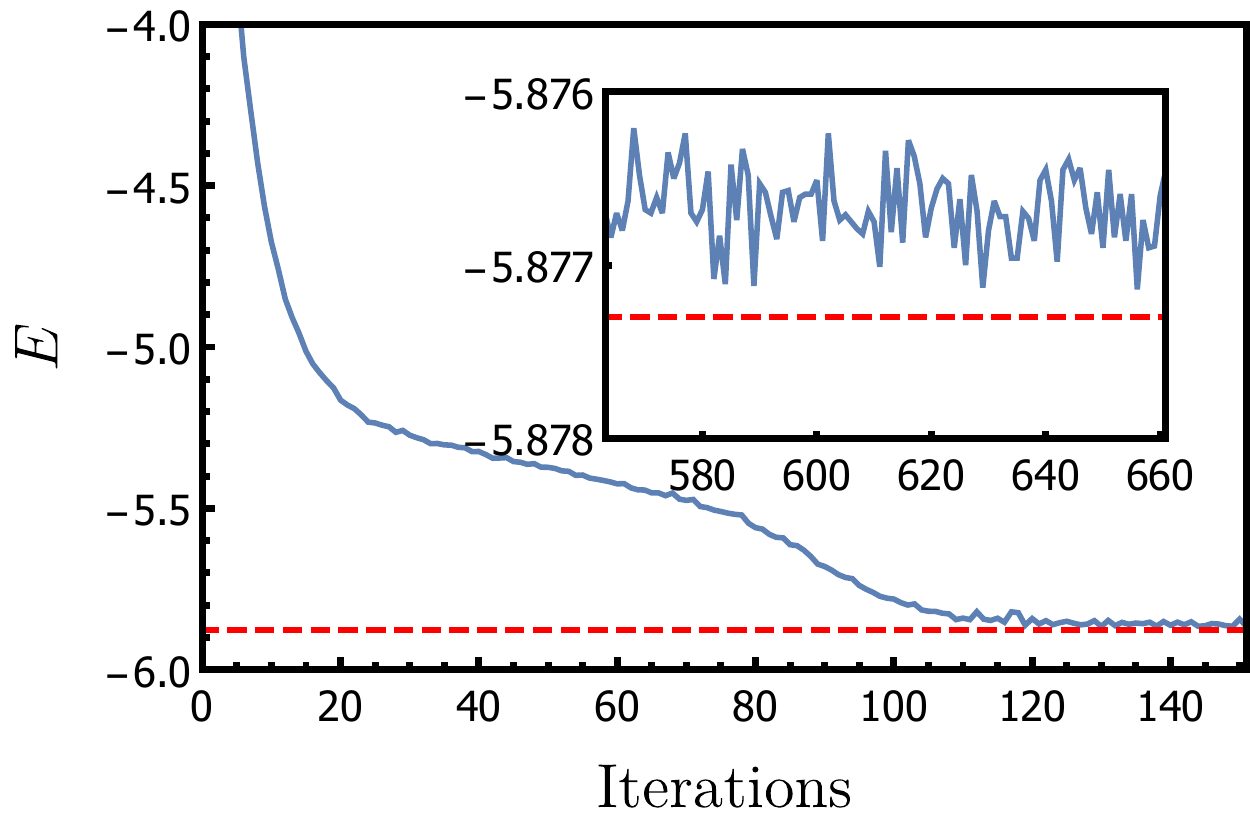}
	\caption{(color online) Energy expectation value $E=\left\langle\psi_{\mathcal{W}}|H|\psi_{\mathcal{W}}\right\rangle/\left\langle\psi_{\mathcal{W}}|\psi_{\mathcal{W}}\right\rangle$ as a function of the stochastic reconfiguration (SR) iterations, for system size $6 \times 4$. For this system size, the expectation values are calculated using a sample of $10000$ configurations drawn with a standard Metropolis algorithm. The inset shows the final SR iterations,  and the horizontal red line marks the exact ground state energy from ED.}
	\label{fig:convergence}
\end{figure}

In Fig.~\ref{fig:convergence}, we show an example of the variational energy of the RBM wavefunction towards the exact ground state energy (red horizontal line) as a function of the number of SR iterations, for a cylinder of size $L_x=6, L_y=4$. 

\section{Cluster neural network quantum states and chiral topological phases } 
\label{sec:cnqs}
To gain analytical insight in how ANN states can exactly represent model wavefunctions for CTPs, we now construct a modified RBM architecture coined cluster neural network states (CNQS). To this end, we associate a fixed number $m$ of auxiliary variables to every subset of $n$ physical spins, coined a cluster of size $n$. We illustrate our construction for the case $n=2, m=1$, where an auxiliary variable $a_{ij} \in \left\{-1, 1\right\}$ is associated with every bond between two distinct physical sites (spins) $i \ne j$ [see Fig.~\ref{fig:two}] to which it is coupled by the complex weights $w_{ij},\tilde w_{ij} \in \mathcal W$. 
The network energy of this RBM is then defined as $\mathcal{E}_{\text{nw}}(\mathcal S, \mathcal W, \mathcal A)=\sum_{i<j}(w_{ij}\sigma_i + \tilde w_{ij} \sigma_j+b_{ij})a_{ij}$, where the $b_{ij}\in \mathcal W$ are the complex local fields for the auxiliary variables. The explicit form of the variational wavefunction at fixed couplings $\mathcal W$ then reads as
\begin{align}
\psi_{\mathcal W}(\mathcal S)&=\sum_{\left\{a_{ij}\right\}}\text{e}^{-\mathcal{E}_{\text{nw}}(\mathcal S,\mathcal W,\mathcal A)}\nonumber\\
&=\prod_{i<j}2\cosh(w_{ij}\sigma_i + \tilde w_{ij} \sigma_j+b_{ij}).
\label{eqn:generalpsi2}
\end{align}
The generalization of this CNQS to larger $n,m$ is straightforward with the number of couplings in $\mathcal W$ as well as of the auxiliary variables in $\mathcal A$ scaling as $m N^n$. The generalization of the product structure of $\psi_{\mathcal W}$ in Eq.~(\ref{eqn:generalpsi2}) then contains $m$ factors for each cluster labeled by $n$ indices $i_1<i_2\ldots < i_n$, capturing $n$-body correlations.

\emph{Chiral topological phases from CNQS. } As a concrete example, we now demonstrate how the above CNQS construction can be used to exactly represent chiral topological states. As a paradigmatic example, we explicitly parametrize a chiral spin liquid ground state of a spin 1/2 system, or equivalently the bosonic $\nu=1/2$ bosonic Laughlin state in the language of hardcore bosons. The desired state $\lvert\psi_L\rangle$ in the complex position representation $z_j=x_j+iy_j$ is written as
\begin{align}
\psi_L(z_1,\ldots,z_p)= \prod_{i<j=1}^p(z_i-z_j)^2\text{e}^{-\frac{\lvert z_i\rvert^2+\lvert z_j\rvert^2}{p-1}},
\label{eqn:Laughlin}
\end{align}
where $p$ is the number of particles. In our spin 1/2 representation, where we choose $\sigma_i\in \left\{0,1\right\}$, the positions of the up-spins, i.e. sites with $\sigma_i=1$ are simply identified with the positions $z_i$ of hardcore bosons. In order to represent $\lvert\psi_L\rangle$ as a CNQS, it is helpful to rewrite Eq.~(\ref{eqn:Laughlin}) as
\begin{align}
\psi_L(\mathcal S)= \prod_{i<j=1}^N\left[1+\left((z_i-z_j)^2\text{e}^{-\frac{\lvert z_i\rvert^2+\lvert z_j\rvert^2}{p-1}}-1\right)\sigma_i\sigma_j\right],
\label{eqn:LaughlinCNQS}
\end{align}
where in the CNQS language only pairs $i,j$ ($2$-clusters) with both sites occupied ($\sigma_i=\sigma_j=1$) contribute a non-trivial factor to the wavefunction. Eq.~(\ref{eqn:LaughlinCNQS}) is of the general Jastrow form $\prod_{i<j=1}^N(c_{ij} \sigma_i +d_{ij} \sigma_j + e_{ij}\sigma_i\sigma_j+f_{ij})$ with arbitrary complex coefficients $c,d,e,f$. Simple parameter counting shows that any such state can be \emph{exactly} represented as a CNQS with $n=m=2$. This already tells us that, the exact Laughlin wavefunction $\psi_L$ is part of the variational space for $n=m=2$. 

Going beyond this general argument, we analytically find that even with $m=1$ and $w_{ij}=\tilde w_{ij}=-2b_{ij}$, i.e. with a single complex parameter per $ij$-pair, the $ij$-factor of the CNQS wavefunction (\ref{eqn:generalpsi2}) can be decomposed as
\begin{align}
&\cosh(w_{ij}\sigma_i + \tilde w_{ij} \sigma_j+b_{ij})=\nonumber\\
&\cosh(b_{ij})\left(1+\left[\frac{\cosh(3b_{ij})}{\cosh(b_{ij})}-1\right]\sigma_i\sigma_j\right).
\label{eqn:singlepar}
\end{align}
Comparing Eq.~(\ref{eqn:singlepar}) to Eq.~(\ref{eqn:LaughlinCNQS}), we find that any Laughlin wavefunction up to a global prefactor can be exactly represented with analytically determined parameters $b_{ij}$.

\section{Concluding discussion } 
\label{sec:discussion}
Using ANN constructions for variational quantum many-body wave functions has already led to several promising insights, including the parameterization of states with volume law entanglement \cite{DengVolumelaw}, the approximate representation of $p+ip$ superconductors \cite{HuangPiP}, the exact representation of topological stabilizer states \cite{RBMStabilizer}, a numerical study of the 2D-Hubbard model \cite{SaitoHubbard, NomuraHubbard}, and on the relation between ANN states and conventional tensor networks \cite{ChenANN_TN}. 

Here, we have shown that RBM states can be efficiently used as an ansatz to describe chiral topological phases, both at the numerical level and at an exact analytical level. With small-scale numerical benchmark studies not imposing any symmetry constraints except particle number conservation, we could already significantly exceed the system sizes amenable to direct study with exact diagonalization. However, due to the
expected polynomial cost of our RBM simulations \cite{footPoly}, even larger systems sizes should be tractable. This may be of particular importance for gapless topological phases exhibiting severe finite size effects \cite{CriticalCTP}. Moreover, as generally shown in Ref. \cite{CarleoTroyer}, the ANN approach is capable of describing unitary time-evolution. This may open up the possibility to study dynamical aspects such as non-equilibrium response functions and quantum transport properties of CTPs, where comparably large system sizes are required to clearly observe topologically quantized features, and where capturing quantum correlations beyond area law entanglement is important.

The fact that certain CTP model states can naturally be parameterized with polynomial cost within the ANN approach is generally promising, as their exact parameterization with the most well known tensor network methods such as matrix product states requires exponential cost in at least one spatial direction \cite{Zaletel2012}. However, it remains an open question whether the fundamental limitation \cite{Read2017} to the representability of non-trivial CTPs with tensor network states using finite resources in the thermodynamic limit can be overcome with ANN states. An important challenge and interesting direction of future research hence is to devise ANN architectures that are flexible enough to parameterize even in the thermodynamic limit CTPs and other strongly correlated topological phases with no known exact tensor network representative.

\section*{Acknowledgements. } We acknowledge discussions with  E. Bergholtz, G. Carleo, M. Heyl, M. Hohenadler, N. Cooper, C. Repellin, and A. Sterdyniak. The numerical calculations were performed on resources at the Chalmers
Centre for Computational Science and Engineering (C3SE) provided by the Swedish National Infrastructure for Computing (SNIC). We acknowledge financial support from the German Research Foundation (DFG) through the Collaborative Research Centre SFB 1143.\\

\emph{Note Added. } While preparing this manuscript for submission, two related preprints appeared on the arXiv \cite{GlasserMunich,Clark}. S. R. Clark constructs a mapping \cite{Clark} between RBM states and correlator product states, with relevance for CTP states such as Laughlin wavefunctions. I. Glasser et al. \cite{GlasserMunich} establish a correspondence between string-bond network states and RBM states, also presenting VMC data for the $\nu=1/2$ Laughlin phase, but studying a different model Hamiltonian \cite{NielsenCSL} from the one in our manuscript.

\section*{Appendix: }

\subsection*{A1. Stochastic Reconfiguration}
In the following we provide a brief description of the stochastic reconfiguration (SR) method \cite{Sorella2001,SorellaSRchem1,SorellaSR}. The problem the SR method addresses is the minimization of the energy expectation value within the subspace of the variational wavefunctions. In order to carry out this minimization procedure we interpret the variational state as effectively depending on $2N_{\text{w}}$ real parameters, which are the real and imaginary parts of the $N_{\text{w}}$ complex weights. We denote with the real vector $\boldsymbol{w}$ a certain configuration of real and imaginary parts of the weights, and with $|\hat{\psi}_{\boldsymbol{w}}\rangle=\frac{|\psi_{\boldsymbol{w}}\rangle}{\sqrt{\langle\psi_{\boldsymbol{w}}|\psi_{\boldsymbol{w}}\rangle}}$ the normalized variational state for this set of values. We adopt the following convention: $w_j$ for $j=2\ell-1$ is the real part of the $\ell\,$th complex weight, and for $j=2\ell$ it is the imaginary part of the $\ell\,$th complex weight, where $\ell=1,\dots,N_{\text{w}}$. In the VMC algorithm, after the samples from the probability distribution $\langle\hat{\psi}_{\boldsymbol{w}}|\hat{\psi}_{\boldsymbol{w}}\rangle$ have been generated and the energy expectation value $E_{\boldsymbol{w}}=\langle\hat{\psi}_{\boldsymbol{w}}|H|\hat{\psi}_{\boldsymbol{w}}\rangle$ has been calculated, an updating step $d\boldsymbol{w}$ in parameter space is made such that $E_{\boldsymbol{w}+d\boldsymbol{w}}$ is lowered. The SR method ensures an optimal direction of $d\boldsymbol{w}$ by effectively implementing an imaginary time evolution projected onto the variational manifold \cite{CarleoTroyer}. In the following, we discuss the practical implementation of this method to first order in the imaginary time step $d\tau$, as used in our present simulations.\\
Let us introduce the local tangent space $\mathcal{T}_{\boldsymbol{w}}$ at point $\boldsymbol{w}$ to the manifold of variational states parametrized by the weights $w_k$ ($k=1,\ldots,2N_{\text{w}}$). $\mathcal{T}_{\boldsymbol{w}}$ is spanned by the non-orthogonal basis states:
\begin{equation}
\ket{j_{\boldsymbol{w}}}\,=\,|\partial_{w_j}\hat{\psi}_{\boldsymbol{w}}\rangle-|\hat{\psi}_{\boldsymbol{w}}\rangle\langle\hat{\psi}_{\boldsymbol{w}}|\partial_{w_j}\hat{\psi}_{\boldsymbol{w}}\rangle \,\,.
\end{equation}
Notice that $\langle\hat{\psi}_{\boldsymbol{w}}|j_{\boldsymbol{w}}\rangle=0\quad\forall\;j=1,\ldots,2N_{\text{w}}$. We again point out that the derivatives $\partial_{w_j}$ are derivatives with respect to real parts for odd $j=2\ell-1$, and with respect to the imaginary parts of the complex weight $\ell$ for even $j=2\ell$. We denote with $(S_{\boldsymbol{w}})_{j,k}=\langle j_{\boldsymbol{w}}|k_{\boldsymbol{w}}\rangle$ the components of the local metric tensor at $\boldsymbol{w}$, also referred to as the covariance matrix, which take the form
\begin{equation}
(S_{\boldsymbol{w}})_{j,k}= \langle\partial_{w_j}\hat{\psi}_{\boldsymbol{w}}|\partial_{w_k}\hat{\psi}_{\boldsymbol{w}}\rangle-\langle\partial_{w_j}\hat{\psi}_{\boldsymbol{w}}|\hat{\psi}_{\boldsymbol{w}}\rangle\langle\hat{\psi}_{\boldsymbol{w}}|\partial_{w_k}\hat{\psi}_{\boldsymbol{w}}\rangle \,\,.
\label{eq:metrictensorS}
\end{equation}
With $\tau$ being the imaginary time, and assuming that the wavefunction depends on $\tau$ through the variational parameters $w_k(\tau)$, the imaginary time evolution is governed by the equation \begin{equation}
|\hat{\psi}_{\boldsymbol{w}(\tau+d\tau)}\rangle\,=\,e^{-d\tau H}|\hat{\psi}_{\boldsymbol{w}(\tau)}\rangle\,\,.
\end{equation}
Expanding the left-hand side of the above equation to first order in $d\tau$ we obtain
\begin{align*}
|\hat{\psi}_{\boldsymbol{w}(\tau+d\tau)}\rangle\simeq&|\hat{\psi}_{\boldsymbol{w}(\tau)}\rangle+d\tau\sum_{k=1}^{2N_{\text{w}}}\dot{w_k}(\tau)\Big[|\partial_{w_k}\hat{\psi}_{\boldsymbol{w}(\tau)}\rangle\\&-|\hat{\psi}_{\boldsymbol{w}(\tau)}\rangle\langle \hat{\psi}_{\boldsymbol{w}(\tau)}|\partial_{w_k}\hat{\psi}_{\boldsymbol{w}(\tau)}\rangle\Big]
\end{align*}
where the second term in the sum subtracts the variation of the state parallel to $|\hat{\psi}_{\boldsymbol{w}(\tau)}\rangle$ (to keep the norm fixed), and $\dot{w_k}$ denotes the imaginary time derivative of $w_k$. The right-hand side expanded to first order reads as
\begin{equation}
e^{-d\tau H}|\hat{\psi}_{\boldsymbol{w}(\tau)}\rangle\,\simeq\,|\hat{\psi}_{\boldsymbol{w}(\tau)}\rangle-d\tau\,H|\hat{\psi}_{\boldsymbol{w}(\tau)}\rangle\,\,.
\end{equation}
Equating the two terms and multiplying from the left by $\langle j_{\boldsymbol{w}(\tau)}|$ (i.e. projecting the imaginary time evolution onto the tangent space $\mathcal{T}_{\boldsymbol{w}(\tau)}$) we obtain (we drop the $\tau$ dependence now for simplicity)
\begin{align*}
&\sum_{k=1}^{2N_{\text{w}}}\dot{w_k}\Big[\langle\partial_{w_j}\hat{\psi}_{\boldsymbol{w}}|\partial_{w_k}\hat{\psi}_{\boldsymbol{w}}\rangle-\langle\partial_{w_j}\hat{\psi}_{\boldsymbol{w}}|\hat{\psi}_{\boldsymbol{w}}\rangle\langle\hat{\psi}_{\boldsymbol{w}}|\partial_{w_k}\hat{\psi}_{\boldsymbol{w}}\rangle\Big]=\\
&=-\langle\partial_{w_j}\hat{\psi}_{\boldsymbol{w}}|H|\hat{\psi}_{\boldsymbol{w}}\rangle+\langle\partial_{w_j}\hat{\psi}_{\boldsymbol{w}}|\hat{\psi}_{\boldsymbol{w}}\rangle\langle\hat{\psi}_{\boldsymbol{w}}|H|\hat{\psi}_{\boldsymbol{w}}\rangle
\end{align*}
which can be rewritten in vector notation as
\begin{equation}
S_{\boldsymbol{w}}\frac{d\boldsymbol{w}}{d\tau}=-\boldsymbol{F}_{\boldsymbol{w}}
\end{equation}
where $S_{\boldsymbol{w}}$ is the $2N_{\text{w}}\times 2N_{\text{w}}$ metric tensor [see Eq. (\ref{eq:metrictensorS})] and $\boldsymbol{F}_{\boldsymbol{w}}$ is the force vector whose components are given by
\begin{equation}
F_j(\boldsymbol{w})=\langle\partial_{w_j}\hat{\psi}_{\boldsymbol{w}}|H|\hat{\psi}_{\boldsymbol{w}}\rangle-\langle\partial_{w_j}\hat{\psi}_{\boldsymbol{w}}|\hat{\psi}_{\boldsymbol{w}}\rangle\langle\hat{\psi}_{\boldsymbol{w}}|H|\hat{\psi}_{\boldsymbol{w}}\rangle
\label{eq:forcevector}
\end{equation}
Introducing the imaginary time step size $\gamma$ we then have
\begin{equation}
d\boldsymbol{w}=-\,\gamma\,S^{-1}_{\boldsymbol{w}}\boldsymbol{F}_{\boldsymbol{w}}\,\,.
\label{eq:SRstep}
\end{equation}
At each imaginary time step the covariance matrix and the force vector elements are calculated from the samples of $\langle\hat{\psi}_{\boldsymbol{w}}|\hat{\psi}_{\boldsymbol{w}}\rangle$ by computing the local variational derivative estimators \cite{Sorella2001,SorellaSRchem1,SorellaSR,CarleoTroyer}
\begin{equation}
O_k(\mathcal{S})=\frac{\partial}{\partial w_k}\log\left(\langle\mathcal{S}|\psi_{\boldsymbol{w}}\rangle\right)
\label{eq:localDerivEstim}
\end{equation}
at spin configuration $\mathcal{S}$, and using
\begin{align}
&S_{k,k'}(\boldsymbol{w})=\left\langle O_k^*O_{k'}\right\rangle-\left\langle O_k^*\right\rangle\left\langle O_{k'}\right\rangle \,\,,\\
&F_k(\boldsymbol{w})=\left\langle O_k^*E_{\text{loc}}\right\rangle-\left\langle E_{\text{loc}}\right\rangle\left\langle O_k^*\right\rangle
\end{align}
with $E_{\text{loc}}(\mathcal{S})=\frac{\langle\mathcal{S}|H|\psi_{\boldsymbol{w}}\rangle}{\langle\mathcal{S}|\psi_{\boldsymbol{w}}\rangle}$, and the square brackets denoting the Monte Carlo average over the samples.\\
The step $d\boldsymbol{w}$ calculated in Eq. (\ref{eq:SRstep}) is a \emph{complex} vector with $2N_{\text{w}}$ components which correspond to the variations of real and imaginary parts of the $N_{\text{w}}$ complex weights. Denoting with $u_{\ell}$ the $\ell\,$th complex weight, the SR update $du_{\ell}=du_{\ell}^R+i\,du_{\ell}^I$ is calculated from $d\boldsymbol{w}$ as
\begin{equation}
\begin{cases}
du_{\ell}^R=\text{Re}(dw_{2\ell-1})-\text{Im}(dw_{2\ell}) \\ du_{\ell}^I=\text{Re}(dw_{2\ell})+\text{Im}(dw_{2\ell-1}) \,\,\,.
\end{cases}
\label{eq:RealImagComplexWeightUpdate}
\end{equation}

\subsection*{A2. Efficient Calculation of Step in Parameter Space}
Rather than explicitly evaluating the $S$ matrix inverse for calculating the step in parameter space from Eq. (\ref{eq:SRstep}) it is numerically more efficient to solve the linear system
\begin{equation}
S_{\boldsymbol{w}}\,d\boldsymbol{w}=-\,\gamma\,\boldsymbol{F}_{\boldsymbol{w}}\,\,.
\label{eq:SRlinsys}
\end{equation}
for $d\boldsymbol{w}$. We adopt the MINRES-QLP algorithm \cite{MinresQLP}, which is an iterative linear solver based on the Lanczos method. Lanczos tridiagonalization requires the calculation of the Krylov space which involves matrix-vector multiplications of the form $S_{\boldsymbol{w}}\boldsymbol{v}$, where $\boldsymbol{v}$ is a generic vector with $2N_{\text{w}}$ entries. Since the explicit calculation of the $S$ matrix has a computational cost of $O(N_sN_{\text{w}}^2)$ with $N_s$ being the number of samples, we exploit the product structure of the covariance matrix to avoid its explicit calculation \cite{CarleoTroyer}.  At every SR iteration, for each sample $\mathcal{S}_n$ we store the local variational derivative estimator $O_j(\mathcal{S}_n)$ defined in Eq. (\ref{eq:localDerivEstim}) in the $N_s\times 2N_{\text{w}}$ matrix $\mathbb{O}$, with elements $(\mathbb{O})_{nj}=O_j(\mathcal{S}_n)$. After $\mathbb{O}$ has been evaluated we compute
\begin{equation}
\underline{u}=\mathbb{O}\,\boldsymbol{v}
\end{equation}
where $\underline{u}$ is a $N_s$ component vector with elements
\begin{equation}
(\underline{u})_n=\sum_{j=1}^{2N_{\text{w}}}\frac{(\partial_{w_j}\psi_{\boldsymbol{w}})(\mathcal{S}_n)}{\psi_{\boldsymbol{w}}(\mathcal{S}_n)}\,(\boldsymbol{v})_j \,\,.
\end{equation}
Then the evaluation of $\boldsymbol{v}\,'=\frac{1}{N_s}\,\mathbb{O}^{\dagger}\underline{u}$ leads to
\begin{align*}
(\boldsymbol{v}\,')_k&=\sum_{j=1}^{2N_{\text{w}}}\frac{1}{N_s}\sum_{n=1}^{N_s}O_j^*(\mathcal{S}_n)\,O_j(\mathcal{S}_n)\,(\boldsymbol{v})_j \\
&=\sum_{j=1}^{2N_{\text{w}}}\left\langle O^*_kO_j\right\rangle\,(\boldsymbol{v})_j
\end{align*}
and it is sufficient to shift these components by
\begin{equation*}
(\boldsymbol{v}\,')_k\to(\boldsymbol{v}\,')_k-\left\langle O^*_k\right\rangle\sum_{j=1}^{2N_{\text{w}}}\left\langle O_j\right\rangle\,(\boldsymbol{v})_j
\end{equation*}
to retrieve $\boldsymbol{v}\,'=S_{\boldsymbol{w}}\boldsymbol{v}$ at an overall computational cost of $O(N_sN_{\text{w}})$.\\

\subsection*{A3. Metric Rescaling of Step Length}
In our numerical simulations we used a time-dependent imaginary time step $\gamma$, and adopted a Local Metric Rescaling (LMR) technique for the optimization of its length \cite{AmariLMR}, as we explain below. At each imaginary time step, the length of the step in parameter space is rescaled according to the local metric in order to keep the effective step length in the variational manifold constant despite the non-orthogonal frame. Let us consider a generic real function $f_{\boldsymbol{w}}$ which depends on the variational parameters $\boldsymbol{w}$ through the state $|\hat{\psi}_{\boldsymbol{w}}\rangle$, i.e. a real function on the variational manifold embedded in the Hilbert space. This function could be the energy expectation value, the squared modulus of the overlap of $|\hat{\psi}_{\boldsymbol{w}}\rangle$ with a given state, or the distance between $A|\hat{\psi}_{\boldsymbol{w}}\rangle$ and $B|\hat{\psi}_{\boldsymbol{w}}\rangle$ with $A$ and $B$ some operators. Our problem is to find the optimal variation $d\boldsymbol{w}$ of the variational parameters in the context of minimizing $f_{\boldsymbol{w}}$. To this end we Taylor expand to first order
\begin{equation}
f_{\boldsymbol{w}+\epsilon\, d\boldsymbol{w}}\simeq f_{\boldsymbol{w}}+\epsilon\,d\boldsymbol{w}\cdot\boldsymbol{\nabla}_{\boldsymbol{w}}f_{\boldsymbol{w}}
\label{eq:functionTaylor1st}
\end{equation}
where $\epsilon$ is a free small parameter chosen small enough that the above first order approximation is justified. We want to find $d\boldsymbol{w}$ such that $f_{\boldsymbol{w}+\epsilon\,d\boldsymbol{w}}$ is minimal, under the constraint of a fixed step length on the variational manifold, as measured by  $ds_{\boldsymbol{w}+d\boldsymbol{w}}$ for a variation $\boldsymbol{w}\to\boldsymbol{w}+d\boldsymbol{w}$. Explicitly we get
\begin{equation}
ds_{\boldsymbol{w}+d\boldsymbol{w}}=\sqrt{\sum_{i,j}dw_i^*\,(S_{\boldsymbol{w}})_{i,j}\,dw_j}\,=\,1
\end{equation}
where $(S_{\boldsymbol{w}})_{i,j}$ are the components of the metric (or covariance matrix) defined before [see Eq. (\ref{eq:metrictensorS})]. In the following we drop the subscripts in $ds_{\boldsymbol{w}+d\boldsymbol{w}}$ and $S_{\boldsymbol{w}}$ for simplicity. This constrained optimization problem amounts to the minimization of the function
\begin{equation}
\mathcal{L}(d\boldsymbol{w},\lambda)=d\boldsymbol{w}\cdot\boldsymbol{\nabla}_{\boldsymbol{w}}f_{\boldsymbol{w}}+\lambda\big(ds^2-1\big)
\end{equation}
yielding the system
\begin{equation*}
\begin{cases}
\boldsymbol{\nabla}_{d\boldsymbol{w}}\mathcal{L}=\boldsymbol{\nabla}_{\boldsymbol{w}}f_{\boldsymbol{w}}+2\lambda\,S\,d\boldsymbol{w}=0\\
\partial_{\lambda}\mathcal{L}=ds^2-1=0 \,\,.
\end{cases}
\end{equation*}
From the first equation we have
\begin{equation}
d\boldsymbol{w}=-\frac{1}{2\lambda}S^{-1}\boldsymbol{\nabla}_{\boldsymbol{w}}f_{\boldsymbol{w}}\equiv\frac{1}{2\lambda}\delta\boldsymbol{w}
\end{equation}
where we have introduced the \emph{bare} step in parameter space $\delta\boldsymbol{w}=-S^{-1}\boldsymbol{\nabla}_{\boldsymbol{w}}f_{\boldsymbol{w}}$. Plugging this result into the second equation we obtain
\begin{equation}
\lambda=\pm\frac{1}{2}\sqrt{\delta\boldsymbol{w}^{\dag}S\,\delta\boldsymbol{w}} 
\end{equation}
which we call the LMR factor. Since we want the variation of the function $f_{\boldsymbol{w}}$ to be negative we pick the positive root for $\lambda$ and finally arrive at
\begin{equation}
d\boldsymbol{w}=-\frac{S^{-1}\boldsymbol{\nabla}_{\boldsymbol{w}}f}{\sqrt{(S^{-1}\boldsymbol{\nabla}_{\boldsymbol{w}}f)^\dag S(S^{-1}\boldsymbol{\nabla}_{\boldsymbol{w}}f)}} \,\,.
\label{eq:rescaledstep}
\end{equation}
At each SR iteration the \emph{bare} step in parameter space $\delta\boldsymbol{w}$ is calculated and its length is rescaled with the LMR factor $\lambda$ so as to make the effective step length in the variational manifold constant. The rescaled step $d\boldsymbol{w}$ of Eq. (\ref{eq:rescaledstep}) is then multiplied by the free parameter $\epsilon$ in order for the first order expansion of Eq. (\ref{eq:functionTaylor1st}) to be valid. In the simulations we have used the SR method, thus we substitute $\boldsymbol{\nabla}_{\boldsymbol{w}}f_{\boldsymbol{w}}\to\boldsymbol{F}_{\boldsymbol{w}}$ where $\boldsymbol{F}_{\boldsymbol{w}}$ is the force vector defined in Eq. (\ref{eq:forcevector}). The \emph{bare} step then becomes $\delta\boldsymbol{w}=-S^{-1}\boldsymbol{F}_{\boldsymbol{w}}$, and the effective update of the weights is $\epsilon\,d\boldsymbol{w}=\frac{\epsilon\,\delta\boldsymbol{w}}{\sqrt{\delta\boldsymbol{w}^{\dag}S\,\delta\boldsymbol{w}}}=\gamma \delta\boldsymbol{w}$. The $\epsilon$ parameter is generally chosen to be time dependent. We started with an initial value $\epsilon=0.1$ and close to the end we reduced it by a factor $10$ for a more accurate minimum search.\\

\subsection*{A4. Regularization of Metric Tensor}
Finally, we discuss some common issues related to the inversion of the $S$ matrix. The matrix elements of $S$ are calculated as Monte Carlo averages and they are subject to statistical fluctuations. These may lead to very small, not even positive eigenvalues of $S$, which could amplify the fluctuations in the force vector when $S^{-1}\boldsymbol{F}_{\boldsymbol{w}}$ is calculated, leading to numerical instabilities of the SR scheme. One can adopt different regularization schemes to avoid those numerical instabilities \cite{SorellaSR,CarleoTroyer}. One scheme amounts to add a term proportional to the identity matrix, shifting all the diagonal elements of the same amount
\begin{equation}
S^{\text{reg.}}=S+\lambda_{\text{reg.}}I
\label{eq:identityreg}
\end{equation}
and the other one is a rescaling of the diagonal elements
\begin{equation}
(S^{\text{reg.}})_{k,k}=(1+\lambda_{\text{reg.}})\,(S)_{k,k}
\label{eq:diagonalreg}
\end{equation}
We found it useful to adopt the identity regularization of Eq. (\ref{eq:identityreg}) for the first ($\sim 500$) iterations, and then switch to the second scheme (Eq. (\ref{eq:diagonalreg})) towards the end of the simulation. With this choice we found a better stability and more smooth convergence towards the ground state, probably due to the fact that the diagonal regularization does not modify the ratio between the eigenvalues of the $S$ matrix.\\

\bibliographystyle{apsrev}

\begin{thebibliography}{35}
\bibitem{footPoly}
We note that the convergence and thus the overall polynomial cost of the VMC calculations can be challenged by complex energy landscapes with quasi-stable minima in the variational space (see Ref. \cite{footref} for a recent discussion in the context of neural networks).
\end{thebibliography}

\end{document}